\documentstyle[aps,psfig,epsfig,twocolumn]{revtex}
 \newcommand{\eins}{\mbox{$1 \hspace{-1.0mm}  {\bf l}$}}
 \newcommand{\ket}[1]{ | \, #1  \rangle}
 \newcommand{\bra}[1]{ \langle #1 \,  |}
\begin{document} \draft
\title{Characterising a universal cloning machine by maximum-likelihood
estimation}
\author{Massimiliano F. Sacchi}
\address{Optics Section, Blackett Laboratory, 
Imperial College London, London SW7 2BZ, United Kingdom \\
and Dipartimento di Fisica `A. Volta', Universit\`a di Pavia 
and Unit\`a INFM,  via
A. Bassi 6, I-27100 Pavia, Italy}
\date{\today} \maketitle
%%%%%%%%%%%%%%%%%%%%%%%%%%%%%%%%%%%%%%%%%%%%%%%%%%%%%%%%%%%%%%%%%%%%%%%%%%%
\begin{abstract}
We apply a general method for the estimation of completely 
positive maps to the 1-to-2 universal 
covariant cloning machine. The method 
is based on the  maximum-likelihood principle, and makes use of 
random input states, along with random 
projective measurements on the output clones. The downhill
simplex algorithm is applied 
for the maximisation of the likelihood functional. 
\end{abstract}
\vskip 1truecm
\section{Introduction}
Perfect cloning of unknown quantum 
systems is forbidden by the laws of quantum mechanics \cite{wootters}.
However, universal covariant cloning has been proposed \cite{bh},
and has been proved to be optimal in terms of fidelity
\cite{oxibm,wer}. Some relevant applications of 
cloning is eavesdropping in quantum cryptography
\cite{gisin}, and engineering of new kinds of joint measurements
\cite{joint}. Now suppose that someone provides you with a physical device, 
telling you that it works as a universal cloning machine. How can you
check experimentally this statement? You could perform different kinds
of measurement on the output of the device, along with different
preparations of the input states. Then, you could compare the
correlations you have inferred from the outcomes with your theoretical 
predictions. However, the most reliable and general way to proceed
could be entirely reconstructing the completely positive (CP) map
that univocally characterizes a physical device. 
\par Recently, a general method 
to solve this relevant issue has been proposed in Ref. \cite{prl}. 
The method is based on the maximum-likelihood (ML) principle, applied to 
the data obtained by random measurements on the output of the device. 
The ML method 
has been used in the context of phase measurement \cite{phase}, and  
to estimate the density matrix \cite{matrix}, some parameters of
interest in quantum optics \cite{qcmc}, and the CP maps of quantum
communication channels \cite{prl}.   For the problem of CP map
reconstruction \cite{prl}, 
the constraints inherent the maximisation can be imposed by
exploiting the isomorphism between CP map from Hilbert spaces 
${\cal H}$ to ${\cal K}$ and {\em non-negative} operators in the
tensor-product space ${\cal K}\otimes {\cal H}$ \cite{depi,jami,xu}. 
Such  isomorphism has already been useful for the 
study of positive maps \cite{xu} and  to address the problem of
separability of CP maps \cite{sepa}. 
\par In this paper, we apply the general method of Ref. \cite{prl} 
to the 1-to-2 universal covariant cloning machine for 
spin-1/2 systems. In Section II we briefly review 
the ML principle and its use in measuring quantum devices. 
Section III presents 
the calculations needed to explicitly 
construct the likelihood functional to be maximised.  Some numerical
results obtained through a Monte Carlo simulation and the simplex
searching algorithm are then shown to confirm the reliability of the
method. Section IV is devoted to conclusions.
\section{Measuring quantum devices}
The maximum-likelihood principle states  that the best
estimation of unknown 
parameters is given by  the values that are  most likely to produce 
the data one experimenter has observed. Hence, 
this principle involves the maximisation of a 
function of the unknown parameters that is given by 
the theoretical probability of getting the collected data. 
\par Consider a sequence of $K$ independent measurements on the output of a 
physical device acting on quantum states.  
Each measurement is described by the element $F_l(x_l)$ of a POVM, 
where $x_l$ denotes the
outcome at the $l$th measurement, and  $l=1,2,...,K$. Let us denote by
$\rho _l$ the state at the input at the $l$th run. 
The probability of getting the string of outcomes
$\vec x=\{x_1,x_2,...,x_K\}$ is given by 
\begin{eqnarray}
p(\vec x)=
\Pi _{l=1}^K \hbox{Tr}[{\cal E} (\rho _l) F_l (x_l)]\;.\label{prod}
\end{eqnarray}
The best estimate of the 
map $\cal E $ maximizes the logarithm of Eq. (\ref{prod}) 
\begin{eqnarray}
{\cal L} ({\cal E})=\sum _{l=1}^K \log 
\hbox{Tr}[{\cal E} (\rho _l) F_l (x_l)]\;\label{log}
\end{eqnarray}
over the set of completely positive maps. The likelihood function ${\cal
L}( {\cal E})$ is  concave, and in the present case it is
defined on the  convex set of CP maps. Its 
maximum is achieved by a single CP map if the data sample is  
sufficiently large, and the set of measurements is a {\em
quorum} \cite{quorum}. 
\par The constraints to be imposed in the maximisation problem 
are the complete positivity and the trace-preserving property  of the
map $\cal E$. A trace-preserving CP 
map is a linear map from operators in Hilbert space ${\cal H}$ to 
operators in $\cal K$ 
which can be written in the Kraus form \cite{nc}
\begin{eqnarray}
{\cal E}(\rho)=\sum_k A_k\,\rho\,A_k^\dag \;,\label{ss}
\end{eqnarray}
where 
\begin{eqnarray}
\sum_k A_k^\dag A_k = \openone _{\cal H}  \;.
\end{eqnarray}
Let dim(${\cal H})=N$ and dim(${\cal K})=M$, and consider an 
orthonormal basis $\{V_i\}$ for the space of linear
operators on $\cal H$, namely
\begin{eqnarray}
\hbox{Tr}[V_i^\dag V_j]=\delta_{ij}
\;,\label{on1}
\end{eqnarray}
and for any operator $O$
\begin{eqnarray}
O=\sum_{i=1}^{N^2}\hbox{Tr}[V_i^\dag O]\,V_i\;.\label{on2}
\end{eqnarray}
Upon defining the operator \cite{depi,jami}
\begin{eqnarray}
S=\sum_{i=1} ^{N^2} {\cal E}(V_i)\otimes V_i^* \;,\label{s}
\end{eqnarray}
where $*$ denotes complex conjugation, 
one can write for linearity 
\begin{eqnarray}
{\cal E}(\rho)=\hbox{Tr}_{\cal H}
[(\openone _{\cal K}\otimes \rho ^T) \,S] \;,\label{trk}
\end{eqnarray}
where  $T$ is the transposition. 
Notice that \cite{lop}
\begin{eqnarray}
\sum_{i=1}^{N^2}V_i\otimes V_i ^*= |\Psi \rangle \langle \Psi |\;, 
\end{eqnarray}
where $|\Psi \rangle $ is given by the (unnormalized) maximally
entangled state \begin{eqnarray}
|\Psi \rangle =\sum_{n=1}^N |n \rangle \otimes |n \rangle    
\;. 
\end{eqnarray}
Hence, one has also 
\begin{eqnarray}
S={\cal E}\otimes \openone (|\Psi \rangle \langle \Psi |).
\end{eqnarray}
Eqs. (\ref{s}) and (\ref{trk}) establish an isomorphism between linear
maps from ${\cal H}$ to ${\cal K} $ and linear
operators on the tensor-product space ${\cal K} \otimes {\cal H}$. 
Complete positivity and trace-preserving property of 
$\cal E$ imply \cite{nota}
\begin{eqnarray}
S \geq 0 \quad \hbox{and }\quad \hbox{Tr}_{\cal K}[S]=\openone _{\cal
H} \;.\label{sus}
\end{eqnarray}
\par For the construction of the likelihood function 
${\cal L}({\cal E})$ the condition $S\geq 0$ is crucial 
\cite{prl}. Actually, it allows to write
\begin{eqnarray}
S=C^\dag C\;,\label{tt}
\end{eqnarray}
where $C$ is an upper triangular matrix, with positive diagonal
elements \cite{chol}. Similarly, one has for  
the density matrices $\rho_l^T$ and the POVM's $F_l(x_l)$ 
\begin{eqnarray}
\rho^T_l=R^\dag_l R_l \;,\qquad F_l(x_l)=A^\dag _l(x_l)A_l(x_l)\;.\label{deco}
\end{eqnarray}
From Eqs. (\ref{trk}), (\ref{tt})  and  (\ref{deco}), 
the likelihood functional in Eq. (\ref{log}) rewrites 
\begin{eqnarray}
{\cal L}({\cal E})&\equiv &{\cal L}(C)=
 \sum _{l=1}^K \log 
\hbox{Tr}[C^\dag C  (R^\dag_l R_l \otimes 
A^\dag _l(x_l)A_l(x_l))] \nonumber \\& =&
 \sum _{l=1}^K \log \sum_{n,m=1}^{NM}\left| 
\langle \!\langle n |C (R^\dag_l \otimes A^\dag _l (x_l)) |
m \rangle \!\rangle  
\right|^2\;,\label{log2}
\end{eqnarray}
where $\{|n \rangle \!\rangle \}$ denotes an orthonormal basis 
for ${\cal H}\otimes {\cal K}$. 
On one hand, the parameterisation in Eq. (\ref{log2}) implicitly
constrains the complete positivity of the map $\cal E$. On the other, the 
argument of the logarithm is explicitly
positive, thus assuring the stability of 
numerical methods to evaluate ${\cal L}(C)$. 
\par The trace-preserving condition is given in terms of the matrix
$S$ by $\hbox{Tr}_{\cal K}[S]=\openone _{\cal H}$. 
However, the constraint $\hbox{Tr}[S]=N$ which follows from 
$\hbox{Tr}_{\cal K}[S]=\openone _{\cal H}$ 
isolates a closed convex subset of the set of positive 
matrices. Hence, the maximum of the concave likelihood functional still 
remains unique under this looser constraint, and 
one can check {\em a posteriori} that 
the condition $\hbox{Tr}_{\cal K}[S]=\openone_{\cal H}$ is
fulfilled. Using the method of Lagrange multipliers, then one
maximises the effective functional
\begin{eqnarray}
{\tilde {\cal L}}(C)={\cal L}(C) - 
\mu \,\hbox{Tr}[C^\dag C]\;,\label{tilde}
\end{eqnarray}
where ${\cal L}(C)$ is given in Eq. (\ref{log2}), and the value of the 
multiplier $\mu $ can be obtained  as follows. Writing 
$S$ in terms of its eigenvectors as 
$S=\sum_{i} s_i^2 |s_i \rangle \!\rangle  \langle \!\langle s_i|$ , 
the maximum likelihood condition $\partial {\tilde{\cal L}}({\cal
C})/\partial s_i=0$ implies  
\begin{eqnarray}
&&\sum _{l=1}^K \frac{\hbox{Tr}[(\rho _l^T \otimes F_l (x_l))
s_i\, |s_i \rangle \!\rangle  \langle \!\langle s_i|]}
{\hbox{Tr}[(\rho _l^T\otimes F(x_l)) S ]}
\nonumber \\&&=
\mu\, \hbox{Tr}[s_i\, |s_i \rangle \!\rangle  \langle \!\langle s_i|]
\;.\label{mu}\end{eqnarray}
Multiplying by $s_i$ and summing over $i$ gives $\mu =K/N$. 
\section{Characterising the universal cloning machine}
We consider now the problem of estimating the CP map pertaining
to the 1-to-2 universal covariant  cloning machine. 
The map is given by \cite{wer} 
\begin{eqnarray}
{\cal E}(\rho )=\frac{2}{3}\,s_2\left(\rho \otimes 
\eins\right)s_2\;,
\label{tr-2}
\end{eqnarray}
where $s_2$ is the projection operator on the symmetric subspace, 
which is spanned by the set of vectors $\{\ket{s_i}\bra{s_i},
i=0\div2\}$, 
with $\ket{s_0}=\ket{00}$, $\ket{s_1}=1/\sqrt{2}(\ket{01}+
\ket{01})$,  and $\ket{s_2}=\ket{11}$, where $\{\ket{0},\ket{1}\}$
is a basis for each spin 1/2 system. Using the lexicographic 
ordering for the basis of the tensor-product Hilbert space \cite{lex} 
one has 
\begin{eqnarray}
s_2=  \left( 
\begin{array}{cccc}
 1 & 0 & 0 & 0\\
 0 & 1/2 & 1/2 & 0 \\
 0 & 1/2 & 1/2 & 0 \\
 0 & 0 & 0 & 1
\end{array} \right )
\;.\label{smat}
\end{eqnarray}
In the following we label with $A$ and $B,C$ the Hilbert spaces supporting
the input state and the two output copies, respectively.
One can apply Eq. (\ref{s}) to
obtain the corresponding matrix $S $. 
Upon using the operator basis $V_i=\frac {1}{\sqrt 2}\sigma _i$, 
with $\sigma _0= \eins $, and $\sigma _i $ ($i=1,2,3$) denoting the
customary Pauli matrices, one has
\begin{eqnarray}
S &=&\frac 23 \,\frac 12\,
\sum_{i=0}^4 \left[(s_2)^{BC}(\sigma _i^B\otimes \openone
_C)(s_2)^{BC}\right]\otimes \sigma _i^{*A}\;
\nonumber \\& =&
\frac 13\,\left[\openone _A\otimes (s_2)^{BC}\right]\left[
\left( \sum_{i=0}^4 \sigma _i^{*A}\otimes\sigma _i^B\right)\otimes 
\openone
_C \right]\nonumber \\&\times &  
\left[\openone _A\otimes (s_2)^{BC}\right]
\nonumber \\& =&
\frac 23\,\left[\openone _A\otimes (s_2)^{BC}\right]\left[
|\Psi \rangle _{AB}{}_{AB}\langle \Psi|\otimes 
\openone
_C\right]\nonumber \\&\times &  
\left[\openone _A\otimes (s_2)^{BC}\right]\;.
\label{sgamma}
\end{eqnarray}
In the lexicographically ordered basis \cite{lex} of 
${\cal H}_A\otimes {\cal H}_B\otimes {\cal H}_C $ the matrix $S $ writes
\begin{eqnarray}
S=\frac 16 \left( 
\begin{array}{cccccccc}
 4 & 0 & 0 & 0 & 0& 2 & 2 & 0 \\
 0 & 1 & 1 & 0 & 0& 0 & 0 & 2 \\
 0 & 1 & 1 & 0 & 0& 0 & 0 & 2 \\
 0 & 0 & 0 & 0 & 0& 0 & 0 & 0 \\
 0 & 0 & 0 & 0 & 0& 0 & 0 & 0 \\
 2 & 0 & 0 & 0 & 0& 1 & 1 & 0 \\
 2 & 0 & 0 & 0 & 0& 1 & 1 & 0 \\
 0 & 2 & 2 & 0 & 0& 0 & 0 & 4 
\end{array} \right )
\;.\label{6mat}
\end{eqnarray}
Now we want to apply the method presented in the previous section 
to reconstruct the matrix $S$. We use random pure states at the input
of the cloning machine
\begin{eqnarray}
|\psi _l \rangle  &=&\cos(\theta _l/2)|0 \rangle_A +e^{i\phi_l }\sin
(\theta _l /2)|1 \rangle_A 
\;. 
\end{eqnarray}
In matrix notation $\rho _l =|\psi _l \rangle \langle \psi_l|$ we
write 
 \begin{eqnarray}
\rho_l=
\frac 12 (\openone _A + \vec \sigma ^A\cdot \vec n_l)\;,\label{rhol}
\end{eqnarray}
with $\vec n_l\!=\!( \sin\theta _l \cos \phi_l,\sin\theta _l \sin \phi_l,
\cos \theta _l)$.
On  the two output clones we perform independent projective 
measurements along random directions 
$\vec r_l\!=\!( \sin\alpha _l \cos \beta_l,\sin\alpha _l \sin \beta_l,
\cos \alpha _l)$ 
and $\vec t_l\!=\! ( \sin\gamma _l \cos \delta _l,\sin\gamma _l 
\sin \delta_l,\cos \gamma _l)$, 
then using the projector
\begin{eqnarray}
F_l(a_l,b_l)=\frac 14 (\openone _B +a_l \vec \sigma ^B\cdot \vec r_l)\otimes 
(\openone_C +b_l \vec \sigma ^C\cdot \vec t_l)
\;, 
\end{eqnarray}
where $a_l$ and $b_l$ denote the outcomes one has indeed 
obtained (the possible values are $\pm 1$). 
Defining 
\begin{eqnarray}
&&\tilde \alpha _l
 = \frac {\alpha _l} {2} +\pi \frac{a_l -1}{4}\;, \\&&
\tilde \gamma _l
 = \frac {\gamma _l} {2} +\pi \frac{a_l -1}{4}\;, 
\end{eqnarray}
the Cholevsky decomposition for $\rho ^T_l$ and $F_l(a_l,b_l)$ writes as in 
Eqs. (\ref{deco}), 
with 
\begin{eqnarray}
R_l=  \left( 
\begin{array}{cc}
 \cos(\theta _l/2) & e^{i\phi_l}\sin(\theta _l/2) \\
 0 & 0
\end{array} \right )
\; 
\end{eqnarray}
and 
\begin{eqnarray}
A_l(a_l,b_l) &=&  \left( 
\begin{array}{cc}
 \cos\tilde \alpha _l  & e^{-i\beta _l}\sin\tilde \alpha _l 
 \\
 0 & 0
\end{array} \right )
\nonumber \\& \otimes & 
\left( 
\begin{array}{cc}
 \cos\tilde \gamma _l  & e^{-i\delta _l}\sin \tilde \gamma _l \\
 0 & 0
\end{array} \right )
\;. 
\end{eqnarray}
It follows that the matrix $Q_l(a_l,b_l)\equiv 
R_l^\dag \otimes A_l^\dag (a_l,b_l)$ that 
multiplies $C$ in the likelihood function (\ref{log2}) is lower
triangular with just the first column different from zero. 
One has explicitly
\begin{eqnarray}
&&[Q_l(a_l,b_l)]_{11}=\cos(\theta_l /2)
 \cos\tilde \alpha _l 
 \cos\tilde \gamma _l 
\nonumber \\
&&[Q_l(a_l,b_l)]_{21}=e^{i\delta_l}\cos(\theta_l /2)
 \cos\tilde \alpha _l  \sin\tilde \gamma _l \nonumber \\ 
&&[Q_l(a_l,b_l)]_{31}=e^{i\beta_l}\cos(\theta_l /2)
 \sin\tilde \alpha _l 
 \cos\tilde \gamma _l \nonumber \\& & 
[Q_l(a_l,b_l)]_{41}=e^{i(\beta_l+ \delta _l)}\cos(\theta_l /2)
 \sin\tilde \alpha _l 
 \sin\tilde \gamma _l \nonumber \\ 
&&[Q_l(a_l,b_l)]_{51}=e^{-i\phi_l}\sin(\theta_l /2)
 \cos\tilde \alpha _l 
 \cos\tilde \gamma _l    \nonumber \\
&&[Q_l(a_l,b_l)]_{61}=e^{i(\delta_l-\phi_l)}\sin(\theta_l /2)
 \cos\tilde \alpha _l 
 \sin\tilde \gamma _l \nonumber \\
&&[Q_l(a_l,b_l)]_{71}=e^{i(\beta_l-\phi_l)}\sin(\theta_l /2)
 \sin\tilde \alpha _l 
 \cos\tilde \gamma _l \nonumber \\& & 
[Q_l(a_l,b_l)]_{81}=e^{i(\beta_l+ \delta _l-\phi_l)}\sin(\theta_l /2)
 \sin\tilde \alpha _l 
 \sin\tilde \gamma _l 
\;. 
\end{eqnarray}
We have now all the ingredients to construct the likelihood function 
in Eq. (\ref{log2}), which will be a function of the 64 real 
parameters that specify the triangular matrix $C$. 
The problem of the maximisation of ${\cal L}(C)$ enters the realm of
programming and numerical algebra optimisation, where various
techniques 
are known \cite{ciarlet}.  
\par In the following we show the results of a simulation obtained by 
applying the method of downhill simplex \cite{ciarlet,simplex} 
to find the maximum of 
the likelihood functional.  
This method is robust and efficient in case of a relatively small
number of parameters. It has been reliably used in the reconstruction of the 
density matrix of radiation field and spin systems \cite{matrix}, and 
in the characterisation of quantum communication channels for qubits
\cite{prl}. 
\par
\begin{figure}[hbt]
\begin{center}
\epsfxsize=.75 \hsize\leavevmode\epsffile{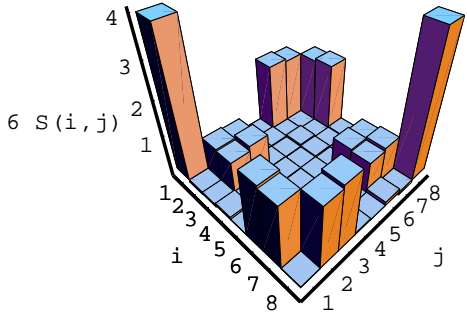}
\end{center}
\caption{Maximum-likelihood reconstruction of the CP map of the 
1-to-2 universal covariant cloning.  The picture represents the values 
of the (real part of the) elements of the matrix $S$.  
Random pure states at the input of the cloning machine, and 
projective measurements 
along random directions on the two clones at the output have been used, with  
$K=10000$ couple of 
measurements. The statistical error in the reconstruction is 
of the order $10^{-2}$. The results compare very well with the
theoretical values of Eq. (\ref{6mat}).} 
\label{f;clon}\end{figure}

 The results are shown in Fig. \ref{f;clon}. Pure states at the input 
of the cloning machine have been used, together with projective 
measurements over the two clones at the
output. In both cases we adopted a uniform distribution on the Bloch
sphere. The Monte Carlo method has been used to generate $K=10000$
data, by using the theoretical probability
\begin{eqnarray}
p(a_l,b_l)=\frac 23 \hbox{Tr}[s_2 (\rho _l \otimes \openone) s_2 \,
F_l(a_l,b_l)]
\;.
\end{eqnarray}
A lengthy but straightforward calculation gives
\begin{eqnarray}
&&p(a_l,b_l)=\frac 14 +\frac 16 (a_l \cos \alpha _l+b_l \cos \gamma
_l)\cos \theta _l \nonumber \\& & 
+\frac 16 [a_l \sin \alpha _l\cos(\beta_l -\phi_l)+b_l \sin \gamma_l
\cos(\delta _l-\phi _l)]\sin \theta _l \nonumber \\& & 
+\frac {1}{ 12}a_l\,b_l\,[\cos \alpha _l\cos \gamma _l+\sin \alpha _l
\sin \gamma _l\cos (\delta _l -\beta _l)]
\;.
\end{eqnarray}
\begin{figure}[hbt]
\begin{center}
\epsfxsize=.75 \hsize\leavevmode\epsffile{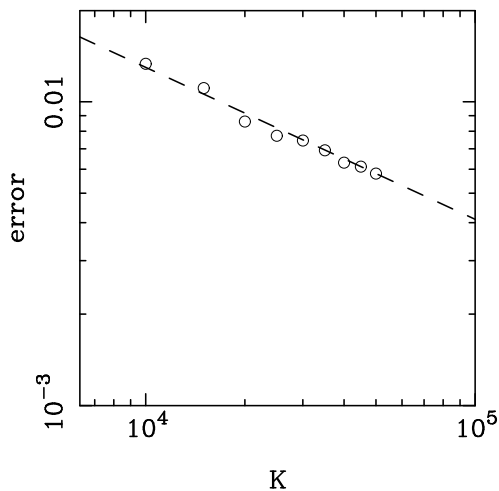}
\end{center}
\caption{Average statistical error in the characterisation of the
universal covariant cloning machine versus number of data $K$. 
The error affects the value of the reconstructed 
elements of the matrix $S$ that is 
univocally related to the CP map of the cloning machine. 
The dotted line represents the asymptotic dependence on the inverse
square root of $K$, in accordance with the central limit theorem.}
\label{f:error}\end{figure}
In Fig. \ref{f:error} we reported the value of the average
statistical error that affects the matrix elements of $S$ versus the
number $K$ of simulated data. In accordance with the central limit theorem 
we find the asymptotic inverse-square-root dependence on $K$. 
\section{Conclusions} 
In conclusion, we have applied a general method to reconstruct
experimentally the completely positive map describing a physical
device to the universal covariant cloning machine. The method, based
on the maximum likelihood principle, involves the maximisation of a
functional, which depends on the results of quantum measurements
performed on the clones at the output. The maximisation has to be made 
over all possible trace-preserving completely positive maps. A suitable 
parametrisation is allowed by the isomorphism between linear map from
Hilbert spaces $\cal H $ to $\cal K$ and linear operators in ${\cal
H}\otimes {\cal K}$, along with the Cholesky decomposition of positive 
matrices. The numerical results we showed here has been obtained by
applying the method of the downhill simplex to search the maximum of
the likelihood functional. In our example, a good characterisation of
the 1-to-2 universal cloning machine has been achieved, with a number
of simulated data as low as $10^4$.  
This is relevant, because some experiments are now feasible, but 
with low data rate or short stability time. In accordance with the central
limit theorem, the statistical error of the characterisation shows the 
inverse-square-root asymptotic dependence on the number of data.  

The method is very general, 
can be implemented immediately in the lab, and can be adopted in 
many fields as quantum optics, spins, optical lattices, 
atoms, ion trap, etc.
\section*{Acknowledgments} The author 
would like to thank the Leverhulme Trust foundation for partial
support. This work has been supported by the Italian Ministero 
dell'Universit\`a e della Ricerca Scientifica e Tecnologica (MURST) 
under the co-sponsored project 1999 {\em Quantum Information
Transmission and Processing: Quantum Teleportation and Error Correction}. 
%%%%%%%%%%%%%%%%%%%%%%%%%%%%%%%%%%%%%%%%%%%%%%%%%%%%%%%%%%

%%%%%%%%%%%%%%%%%%%%%%%%%%%%%%%%%%%%%%%%%%%%%%%%%%%%%%%%%%

\begin{thebibliography}{99}
\bibitem{wootters}W. K.~Wootters and W. H.~Zurek, Nature {\bf 299}, 802
(1982); H. P. Yuen, Phys.~Lett. A {\bf 113}, 405 (1986).
\bibitem{bh} V. Bu\v{z}ek and M.~Hillery, Phys.\ Rev.\ A {\bf 54},
1844 (1996); 
N.~Gisin and S.~Massar, Phys.~Rev.~Lett. {\bf 79}, 2153 (1997). 
\bibitem{oxibm}  D.~Bru\ss , D.~P. DiVincenzo, A.~Ekert, C.~A. Fuchs,
C.~Macchiavello and J.~A. Smolin, Phys. Rev. A {\bf 57}, 2368 (1998); 
D.~Bru\ss , A.~Ekert and C.~Macchiavello, 
Phys. Rev. Lett.  {\bf 81}, 2598 (1998).
\bibitem{wer} R.~F. Werner, Phys.~Rev. A {\bf 58}, 1827 (1998).
\bibitem{gisin} N. Gisin and S. Massar, Phys. Rev. Lett. {\bf 79},
2153 (1997); N. Gisin and B. Huttner, Phys. Lett. A{\bf 228}, 13 (1997). 
\bibitem{joint} G. M. D'Ariano, C. Macchiavello, and M. F. Sacchi, 
quant-ph/0007062; G. M. D'Ariano, F. De Martini, and M. F. Sacchi, 
quant-ph/0012025 (to appear on Phys. Rev. Lett.); 
G. M. D'Ariano and M. F. Sacchi, 
quant-ph/0009080.
\bibitem{prl} M. F. Sacchi, quant-ph/0009104.
\bibitem{phase} S. L. Braunstein, A. S. Lane, and C. M. Caves,
Phys. Rev. Lett. {\bf 69}, 2153 (1992). 
\bibitem{matrix} K. Banaszek, G. M. D'Ariano, M. G. A. Paris,  and M. F. 
Sacchi, Phys.  Rev. A {\bf 61}, 10304(R) (2000).
\bibitem{qcmc} G. M. D'Ariano, M. G. A.  Paris, and M. F. Sacchi, 
Phys. Rev. A {\bf 62}, 023815 (2000); G. M. D'Ariano, M. G. A. Paris, and 
M. F. Sacchi, quant-ph/0009081. 
\bibitem{depi}J. de Pillis, Pacific J. of Math. {\bf 23}, 129 (1967).
\bibitem{jami} A. Jamio\l kowski, Rep. Math. Phys. {\bf 3}, 275
(1972).
\bibitem{xu} S. Yu, Phys. Rev. A {\bf 62}, 024302 (2000).
\bibitem{sepa} J. I. Cirac, W. D\"ur, B. Kraus, and 
M. Lewenstein, quant-ph/0007057.
\bibitem{quorum} For the concept of {\em quorum}, see 
G. M. D'Ariano, L. Maccone, and M. G. A. Paris, quant-ph/0006006; 
Phys. Lett. A {\bf 276}, 25 (2000).
\bibitem{nc} See, for example, M. A. Nielsen and C. M. Caves, 
Phys. Rev. A {\bf 55}, 2547 (1997).
\bibitem{lop} G. M. D'Ariano, P. Lo Presti, and M. F. Sacchi, 
Phys. Lett. A {\bf 272}, 32 (2000).
\bibitem{nota} For a positive---but not completely positive---map 
the condition of positivity of the matrix $S$ is relaxed, by only requiring 
positivity for  tensor product of vectors, namely  
$\langle \phi | \otimes 
\langle \psi |S |\phi \rangle \otimes |\psi \rangle  \geq 0$ 
(see Ref. \cite{jami}).
\bibitem{chol} Such decomposition is referred to as Cholesky
decomposition, and is commonly used in linear programming (see, e.g.,
Ref. \cite{ciarlet}). Moreover, if the matrix $S$ is strictly positive
the decomposition is unique.
\bibitem{lex} We mean that $|i \rangle \otimes |j \rangle  $ 
precedes $|k \rangle \otimes |l \rangle  $ if and only if 
either $i< k$ or $i=k$ and $j<l$.  
\bibitem{ciarlet} See,  for example, P. G. Ciarlet, 
{\em Introduction to numerical 
linear algebra and optimisation}, Cambridge Univ. Press, Cambridge, 1989. 
\bibitem{simplex} W. H. Press, S. A. Teukolsky, W. T. Vetterling, 
B. P. Flannery, 
{\sl Numerical Recipes in Fortran: The Art of Scientific Computing}, 
Cambridge Univ. Press, Cambridge, 1992.
\end{thebibliography}
\end{document}